\documentclass[
unsortedaddress,
amsmath,
amssymb,
aps,
pre,
]{revtex4-2}

\usepackage{graphicx}
\usepackage{dcolumn}
\usepackage{bm}

\usepackage{xcolor}
\usepackage{tikz}

\definecolor{myC0}{rgb}{0.12156862745098039, 0.4666666666666667, 0.7058823529411765}
\definecolor{myC1}{RGB}{255, 127, 14}
\definecolor{mygreen}{rgb}{0., 1.0, 0.}

\newcommand{\uArrowhead}{u_\text{SAR}}

\newcommand{\stressTensorVec}{\boldsymbol{\tau}}
\newcommand{\stressTensor}[1]{\tau_{#1}}

\newcommand{\eigval}[1]{\tau_{#1}}
\newcommand{\eigvecVec}[1]{\mathbf{e}^{(#1)}}

\newcommand{\stressline}[1]{\boldsymbol{\mathcal{T}}_{#1}}

\newcommand{\arclength}{s}

\newcommand{\dirDeriv}[1]{\Tilde{\partial}_{#1}}

\newcommand{\curvature}[1]{\kappa_{#1}}

\newcommand{\forcingVec}{\mathbf{f}}
\newcommand{\forcing}[1]{f_{#1}}

\newcommand{\firstNormalStressDiff}{N_1}

\newcommand{\domainIndexOne}{\text{I}}
\newcommand{\domainIndexTwo}{\text{II}}

\newcommand{\thickness}{\delta}

\newsavebox{\largestimage}

\newcommand{\rev}[1]{\textcolor{black}{#1}}

\begin{document}

\preprint{APS/123-QED}

\title{\textbf{Follow the curvature of viscoelastic stress: \\Insights into the steady arrowhead structure} 
}%

\author{Pierre-Yves Goffin}
\affiliation{Department of Aerospace and Mechanical Engineering, University of Liège, Liège, Belgium}
\author{Yves Dubief}
\affiliation{School of Engineering, University of Vermont, Vermont 05405, USA}
\author{Vincent E. Terrapon}
\email{Contact author: vincent.terrapon@uliege.be}
\affiliation{Department of Aerospace and Mechanical Engineering, University of Liège, Liège, Belgium}

\date{\today}

\begin{abstract}
Focusing on simulated dilute polymer solutions, this letter investigates 
the interactions between flow structures and organized polymer stress sheets for the 
steady arrowhead coherent structure in a two-dimensional periodic channel flow. Formulating the problem in a frame of reference moving with the arrowhead velocity, streamlines, which are also pathlines in this frame, enables the identification of two distinct topological regions linked to two stagnation points. The streamlines help connecting the spatial distribution of polymer stress within the sheets and the dynamics of polymers transported by the flow. Using stresslines,  lines parallel to the eigenvectors of polymer stress, a novel formulation of the viscoelastic stress term in the momentum transport equation proposes a more intuitive interpretation of the relation between the curvature of the stresslines, and the variation of stress along these lines, with the local flow topology. An approximation of this formulation is shown to explain the pressure jump observed in the arrowhead structure as a function of the local curvature of the polymer stress sheet. 

\end{abstract}

\maketitle


\textit{Introduction.} Viscoelastic shear flows obtained from the dilution of polymer additives have the potential to create three phenomena that depart from Newtonian flow dynamics. In supercritical, wall-bounded Reynolds number flows, \rev{polymer} drag reduction (PDR) \cite{white2008mechanics} is the reduction of turbulent friction drag through an interaction with  the dissipation structures generated by near-wall vortices, the engine of wall-bounded turbulence \cite{jimenez1999autonomous}. In wall-free flows, polymers interact similarly with vortices resulting in reduction of turbulence intensity \cite{watanabe2013hybrid,lacassagne2019oscillating}. Under certain polymer concentrations and relaxation time scales, flow structures, distinct from any structures observed in Newtonian flow turbulence, cause the phenomenon of elasto-inertial turbulence (EIT) \cite{dubief2023elasto}, which may exist in subcritical Reynolds number flows and may coexist with inertial (Newtonian) turbulence as in PDR \cite{Dubief2013,sid2018two} in wall-bounded flows and free-shear flows like jets \cite{yamani2023spatiotemporal}. Lastly, polymer additives in inertialess flows can cause chaos, a phenomenon known as elastic turbulence (ET) \cite{steinberg2021elastic}. EIT and ET share a key dynamical feature: viscoelastic stress tends to be organized in thin-sheets \cite{Samanta2013,Dubief2013,terrapon2015role,sid2018two,Dubief2022,Berti2010,kumar2024nested,morozov2022coherent,lellep2024purely}, which have been long speculated to be central to the self-sustaining mechanisms of EIT and ET. \rev{It should be noted that such thin sheets of high polymer stress have also been observed in other contexts, such as viscoelastic laminar flows around obstacles \cite{Harlen1990, Mokhtari}, and have been called birefringent strands owing to their anisotropic optical characteristics.} The dynamical role of these coherent stress sheets in correlation with the kinematic flow structures, \textit{e.g.} regions of low-, high-pressure, recirculation, streaks and vortices, remains poorly understood. 
The objective of the present research is to gain insights into the interactions between viscoelastic stress sheets, and the pressure and velocity coherent structures through an analysis focused on the geometry of the sheets.


The present letter proposes a new approach to study the coherence of the polymer stress or viscoelastic stress tensor in the reference frame of stresslines. Here stresslines are lines tangent to the local eigenvectors of the viscoelastic stress tensor. Within this geometrical transformation, in sheets of high polymer stress, the polymer body force, the term adding the effects of polymers in the transport momentum equations, becomes a function of geometrical properties of the sheets. This approach is applied here to the steady arrowhead regime (SAR) \cite{Dubief2022}. This regime is a traveling wave \rev{made of thin sheets of high polymer stress}, and is one of the attractors of \rev{two-dimensional polymeric channel flows} \cite{Beneitez2023}\rev{. SAR in 2D channel flows} has the particularity of being steady in a frame of reference moving at constant velocity, \rev{thereby greatly facilitating the investigation of the dynamics at play between polymer stress sheets, pressure and velocity \cite{Dubief2022}}. Arrowheads have been observed in several flow configurations (e.g., channel flows \cite{Dubief2022,Page2020}, planar Kolmogorov flows \cite{Berti2010,nichols2025period,Lewy2025}, flows in porous media \cite{zhu2024early}), simulated using different viscoelastic models (e.g., Oldroyd-B \cite{Berti2010,nichols2025period,Lewy2025}, FENE-P \cite{Dubief2022}, PTT \rev{\cite{morozov2022coherent,lellep2024purely})} and spanning a wide range of physical parameters from ET \cite{Berti2010,morozov2022coherent,lellep2024purely} to EIT \cite{Dubief2022}. Arrowheads may thus be considered to be a robust feature of viscoelastic flows. In the present case, it provides a simpler computational experiment, i.e., two-dimensional and steady, to apply the proposed analysis methodology.

The present study seeks to identify the mechanisms that govern the interactions between coherent structures observed in the viscoelastic stress, pressure and velocity fields of an arrowhead flow. We furthermore hypothesize that the arrowhead shares some similarities with interfacial flows, where the interfaces are thin sheets of viscoelastic stress forming the arrowhead. We leverage the proposed transformation to express the jump conditions at the interface.



\textit{Setup.} The steady arrowhead studied here is the solution of the \rev{conservation of mass and momentum for incompressible polymeric flows, using the FENE-P model to simulate the dynamics of polymers and their contribution to the momentum}:
\begin{align}
&\partial_i u_i = 0\,,\label{eq:div}\\
        &\partial_t u_i + u_j \partial_j u_i = -\partial_i p + \frac{\beta}{Re}\partial_{jj}u_i + \frac{1-\beta}{Re}\partial_j\stressTensor{ij} + F\delta_{i1}\,,\label{eq:ui}\\
        &\partial_t C_{ij} + u_k \partial_k C_{ij} = (\partial_k u_i)C_{jk} + C_{ik}(\partial_k u_j ) -\underbrace{\frac{1}{Wi}\left(\frac{C_{ij}}{1-C_{kk}/L^2} - \delta_{ij}\right)}_{:=\stressTensor{ij}} +\frac{1}{ReSc}\partial_{kk}C_{ij}\,,\label{eq:Cij}
\end{align}
where $u_i$ is the velocity vector, $p$ the pressure, $\tau_{ij}$ the polymer stress tensor and $C_{ij}$ the polymer conformation tensor. The equations are \rev{made} non-dimensional using the mean streamwise (bulk) velocity $U_b$, the channel half-height $h$ \rev{and the density $\rho$}. The bulk Reynolds number $Re$ is based on the \rev{zero-shear viscosity $\mu$ of the solution, $Re=\rho U_bh/\mu$}. The Weissenberg number $Wi$ is the ratio between the polymer relaxation time and the characteristic bulk flow time scale $h/U_b$. The parameter $\beta$ corresponds to the ratio of the solvent \rev{viscosity to the zero-shear viscosity of the solution}. The extension parameter $L$ represents the maximum achievable polymer extension normalized by the equilibrium root mean square \rev{extension of a Hookean dumbbell}. The Schmidt number $Sc$ is the ratio between the solution viscosity and the polymer center-of-mass diffusivity. Finally, $F$ is a spatially uniform streamwise forcing term used to drive the flow at constant mass flow rate.

Leveraging the symmetry of the steady two-dimensional arrowhead, only the upper half of the channel is simulated. The streamwise length of the computational domain is $2\pi h$. The boundary conditions are periodic in the streamwise direction $x$. In the wall-normal direction $y$, symmetry and no-slip conditions are applied at $y=0$ and $y=h=1$, respectively. At the wall, $C_{ij}$ is the solution of Eq.~\eqref{eq:Cij} without the advection and diffusion terms. The chosen parameters are $Re=1000$, $Wi=50$, $\beta=0.9$ and $L=90$, representative of the inertial steady arrowhead regime in a dilute polymer solution. Moreover, a reasonably high Schmidt number, $Sc=500$, is used to best approximate the low center-of-mass diffusivity of polymers. 

The system of equations~(\ref{eq:div}--\ref{eq:Cij}) is solved with the open-source pseudo-spectral code \textsc{Dedalus} (version 2) \cite{Burns2020}, using $4096$ Fourier modes in the streamwise periodic direction and $1024$ Chebyshev modes in the wall-normal direction. The Chebyshev basis functions allow for a natural quadratic mesh refinement near the wall and  centerline. This high spectral resolution ensures a virtually zero spatial discretization error (see Sec.~I of the Supplemental Material \cite{supplemental}). A 3rd-order 4-stage implicit-explicit Runge-Kutta scheme is used to integrate the solution in time with a fixed \rev{non-dimensional} time step of $5 \times 10^{-4}$. The combination of the chosen resolution, time step and $Sc$ ensures the positive definiteness of $C_{ij}$ everywhere in the computational domain. Since the arrowhead structure is a traveling wave, the problem is solved in a frame of reference moving with the velocity of the arrowhead. This velocity is iteratively converged based on the solution at the previous time step \rev{and is found to be $\uArrowhead = 1.4578$}.  
The normalized L$_2$-norm of the rate of change of the different transported variables at convergence for the investigated simulation is $\mathcal{O}(10^{-9})$ or smaller, demonstrating that steady-state has been reached (see Sec.~I of the Supplemental Material \cite{supplemental}).  


\begin{figure*}
    \centering
    \includegraphics[width=\textwidth]{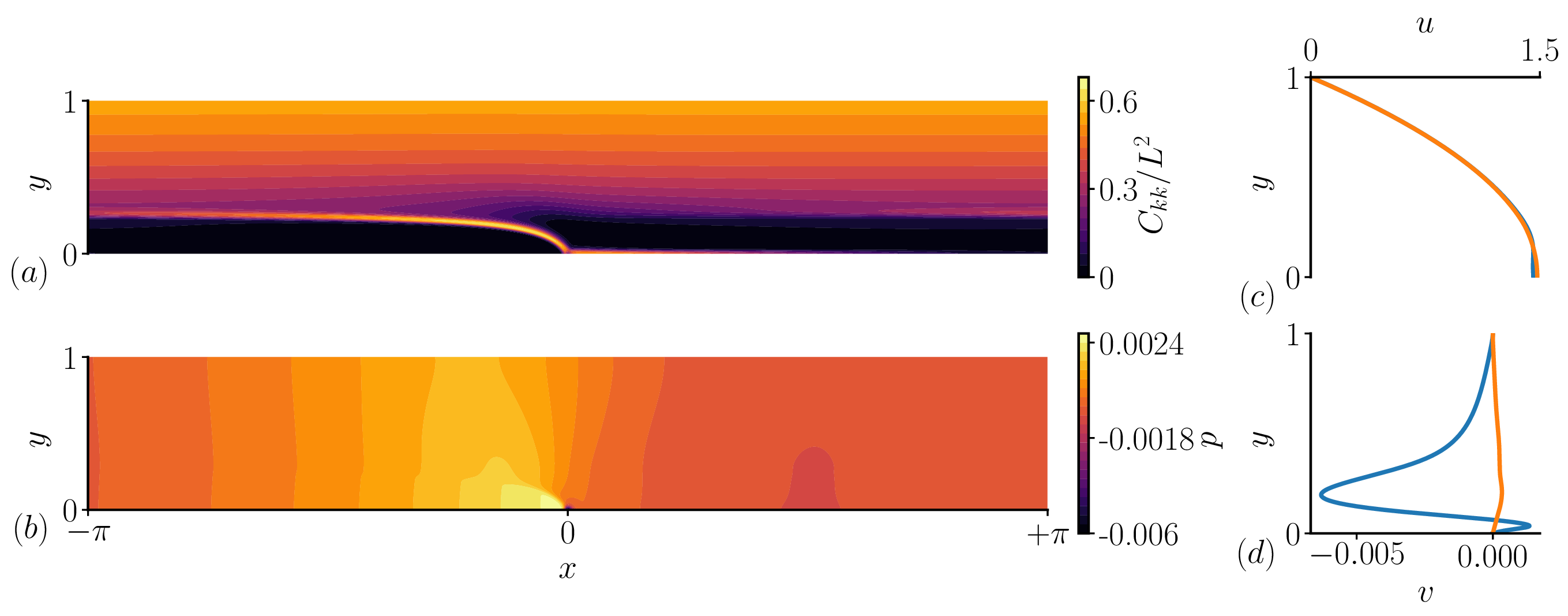}
    \caption{\label{fig:ckk_p_u} Contour plot of the relative polymer extension $C_{kk}/L^2$ (a) and pressure field $p$ (b). Streamwise (c) and wall-normal (d) velocity profiles evaluated at the location of the pressure minimum ($x=0$) (blue) and at $x=\pm\pi$ (orange). Because of the solution symmetry, only the upper half of the channel is shown. \rev{In the chosen moving frame of reference, the flow goes from right to left. $Re=1000$, $Wi=50$, $\beta=0.9$, $L=90$ and $Sc=500$.}}
\end{figure*}

\begin{figure*}
    \includegraphics[width=\textwidth]{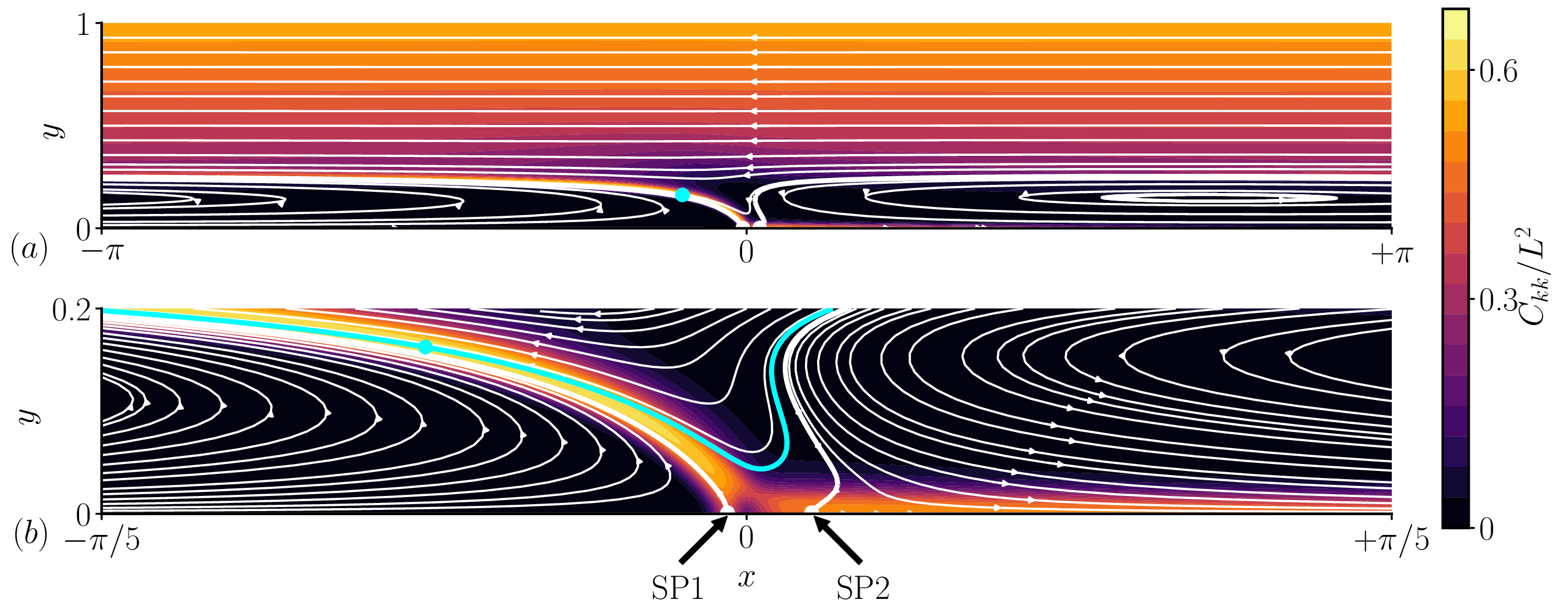}
    \caption{\label{fig:streamlines}Streamlines in the frame of the steady arrowhead overlaid on the contour of the relative polymer extension for the entire computational domain (a) and in the region of the arrowhead (b). The separating streamline (thicker white line) shows the transition between the two topological regions: external flow (a) and the recirculation zone (b). Its end-points on the centerline (white dots) correspond to stagnation points (zero relative velocity); the left stagnation point is denoted SP1 in the following, the other is denoted SP2. The blue dot indicates the position of the maximum of polymer extension, and the blue line is the corresponding streamline. \rev{$Re=1000$, $Wi=50$, $\beta=0.9$, $L=90$ and $Sc=500$.}}
\end{figure*}

\textit{Solution overview.} Figure~\ref{fig:ckk_p_u} shows the polymer extension $C_{kk}/L^2$ (a) and associated pressure fields (b). Two thin-sheets of large polymer stretch are observed: a straight spike-looking sheet along the centerline for $x>0$ and a curved sheet originating from the centerline and moving first away from it before bending downstream until it becomes mostly horizontal and parallel to the wall for $x<0$. The latter is the main structure of the arrowhead.  
In the fraction of the domain where the arrowhead lives, roughly $y\lesssim 1/4$, the polymer stretch is negligible outside of the two sheets. The curved part of the arrowhead produces a significant pressure gradient, as further explained below. Another significant characteristic of the solution is the large and localized pressure minimum at the arrowhead tip, corresponding to the point where the two high polymer stretch sheets join. This low pressure region is also associated with a minimum in polymer extension. The polymer field significantly deviates from the laminar solution (a monotonic increase of polymer stretch from the centerline to the wall, not shown) whereas this particular arrowhead weakly affects the velocity as shown in Fig.~\ref{fig:ckk_p_u}(c-d) for the streamwise, and wall normal velocity profile at the tip of the arrowhead, $x=0$, and towards the tail, $x=\pm\pi$. Augmenting $Wi$ or $L$ increases the impact of the arrowhead on the velocity and pressure fields \cite{Dubief2022}, but at the cost of higher resolution requirements, and without significant qualitative changes in the coherent structures of polymer stress and, pressure or velocity fields. 

Fig.~\ref{fig:streamlines}(a) highlights two distinct regions in the flow topology using streamlines in the moving frame of reference of the arrowhead: (i) a recirculation underneath the arrowhead and above the spike, and (ii) the external flow, between the arrowhead and the wall. The streamlines in the latter region are mostly parallel to the wall, except in the region upstream of the head of the arrowhead and the back of the recirculation. The streamline (thick white line) joining two stagnation points labeled SP1 and SP2 (white dots) defines the boundary between the two regions. Note that SP1 is located at the base of the sheet forming the arrowhead, SP2 is in the spike, upstream of the arrowhead sheet.

\begin{figure*}
    \centering
    \includegraphics[width=\textwidth]{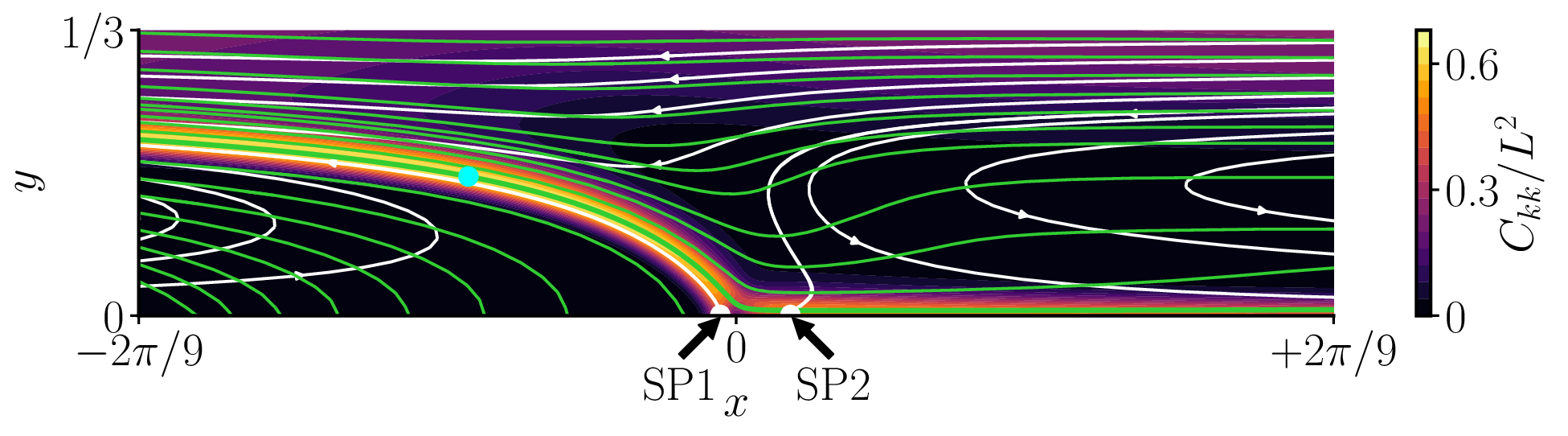}
    \caption{\label{fig:stagnation_points}Illustration of the flow and polymer topology around the two stagnation points. The background contour represents the normalized polymer extension on top of which are shown the streamlines of relative velocity (white), the dividing streamline (thicker white line) and the stresslines of principal stress $\stressline{1}$ (green). The complementary stresslines $\stressline{2}$ are not shown but are everywhere normal to the green lines. The blue dot indicates the position of the overall maximum polymer extension and the two white dots on the centerline represent the two stagnation points. \rev{$Re=1000$, $Wi=50$, $\beta=0.9$, $L=90$ and $Sc=500$.}}
\end{figure*}


\textit{Flow-polymer interaction.} To study the interactions between polymer stress and flow, we first consider how the flow topology can explain the spatial organization of polymer stretching, with a focus on the thin sheets of large polymer extension. The characteristic shape of the sheets illustrated in Fig.~\ref{fig:streamlines} is directly related to extensional flows originating from the two stagnation points. At the spike stagnation point (SP2), the flow moves towards the centerline from above (compression in the $\mathbf{e}_y$-direction) and accelerates away from SP2 along the centerline (extension in the $\mathbf{e}_x$-direction). This horizontal extensional flow stretches the polymers that are then mostly aligned parallel to the centerline, creating the spike in front of the arrowhead. \rev{Far away on the right of SP2 the stresses relax as the deformation rates drop below $Wi~\approx 1$, akin to birefringent strands observed behind a cylinder \cite{Harlen1990,Mokhtari}.} 
The fluid moving away from SP2 towards the left also experiences initially an extensional flow but is then compressed due to a strong deceleration when approaching SP1 from the right. The reason comes from the alignment of the extensional flow originating from SP1 in the $\mathbf{e}_y$-direction, i.e. the flow at SP1 moves away from the centerline in the vertical direction. The rapid compression explains the low polymer extension just right of SP1 associated with the region of low pressure. On the other hand, the flow accelerates along the dividing streamline connecting SP1 and SP2. The response of the polymers to this extensional flow is thus a rapid stretching. However, the streamlines, and thus the sheet, are bent by the external flow and strong shear at the wall, inducing the typical arrowhead shape. Further downstream, the extension rate slowly vanishes and only the contribution of the background shear remains\rev{. It causes the polymer stress sheet to relax to the surrounding stress.} 

At first, one could expect the maximum polymer extension to occur at the stagnation points, as polymers stay in these regions for a large amount of time. Despite a significant extension there, the maximum is actually found further downstream along the curved sheet (blue dot in Fig.~\ref{fig:streamlines}), on a streamline slightly above the dividing streamline. 
Two mechanisms could qualitatively explain this observation. First, the stretching rate is not maximum at the stagnation point but increases along the sheet to reach a maximum further downstream. The flow not only accelerates vertically due to the extensional flow associated with SP1, but also accelerates horizontally when moving away from the centerline due to the background shear (in the frame of reference moving with the arrowhead, the velocity is maximum at the wall). This combined acceleration (illustrated by the contraction of the streamlines approaching the point of maximum extension) induces a strong stretching of the polymer molecules that then decreases as the sheet becomes more horizontal. Additionally, the extended polymer molecules rotate to align with the sheet when moving downstream, but they are never perfectly aligned with it. This misalignment combined with a shear across the sheet amplifies the effect of this shear such that the polymers transiently reach a larger extension than they would in an otherwise constant shear of the same magnitude. While both mechanisms are likely to play a role here, it is difficult to determine which one dominates. Simulations (not shown here) seem to indicate that their relative importance depends on the simulation parameters considered. 

\textit{Polymer-flow interaction.} We now turn our attention to the investigation of how the polymer body force $\forcing{i} = (1-\beta)/Re\,\partial_{j}\stressTensor{ij}$ causes the observed flow topology. We propose to formulate this body force in a system of coordinates associated with the \rev{principal} axes of the stress tensor, which provides a more intuitive interpretation of this body force. Note that we focus here only on the two-dimensional case (see Sec.~II of the Supplemental Material \cite{supplemental} for the extension to the three-dimensional case and reference \cite{Carmo_differential_geometry} therein).

The polymer stress tensor $\stressTensorVec$ may be written in general as
\begin{equation}
    \rev{\stressTensorVec=\stressTensor{ij} = \sum_{\alpha=1}^2 \eigval{\alpha} \eigvecVec{\alpha} \otimes\eigvecVec{\alpha}}\,,
\end{equation}
where $\eigval{\alpha}$ and $\eigvecVec{\alpha}$ are the $\alpha$-th eigenvalue and associated unit eigenvector (index summation rule does not apply to index $\alpha$). Since the tensor $\stressTensorVec$ is symmetric, the eigenvalues are real and the eigenvectors are orthogonal to each other. Assuming a sufficiently smooth $\stressTensorVec(\mathbf{x})$-field, it is possible to construct two families of continuous smooth curves as integral curves of the associated eigenvectors $\eigvecVec{\alpha}$. These so-called ``stresslines'' are defined by the position vectors $\stressline{\alpha}(\arclength)$,  where $\arclength$ is the arc-length of the curve, such that $\text{d}\stressline{\alpha} / \text{d}\arclength = \eigvecVec{\alpha}$.
The two families of curves are normal to each other and can be seen as the equivalent to classical streamlines in the sense that they are everywhere parallel to their associated unit eigenvectors. 
To avoid any ambiguity, we use the common convention that $\alpha=1$ corresponds to the largest eigenvalue, i.e. $\eigval{1} \geq \eigval{2}$. Additionally, we choose the orientation of the eigenvector of one family of stresslines such that it points towards the center of curvature of the other family of stresslines (e.g., $\eigvecVec{1}$ points towards the center of curvature of $\stressline{2}$).

We can now express the polymer body force in the more natural system of coordinates defined by the local principal directions of the stress tensor. After some developments (see Sec.~II of the Supplemental Material \cite{supplemental}) the two components  $\forcing{1}$ and $\forcing{2}$ of the body force, respectively related to the first and second eigenvectors, are given by
\begin{equation}\label{eq:forcing}
    \frac{1-\beta}{Re}\nabla \cdot \stressTensorVec = \forcingVec = \forcing{1}\eigvecVec{1} + \forcing{2}\eigvecVec{2}\,,
    \quad \text{ with }\quad
    \left\{
    \begin{aligned}
    &\forcing{1} = \frac{1-\beta}{Re}\left(\dirDeriv{1}\eigval{1} - \curvature{2}\firstNormalStressDiff\right)\,,\\
    &\forcing{2} = \frac{1-\beta}{Re}\left(\dirDeriv{2}\eigval{2} + \curvature{1}\firstNormalStressDiff\right)\,,
    \end{aligned}
    \right.
\end{equation}
where $\dirDeriv{\alpha}$ is the directional derivative in the $\alpha$-th principal direction, $\curvature{\alpha}$ the local curvature of the stressline $\stressline{\alpha}$ and $\firstNormalStressDiff = \eigval{1} - \eigval{2} \geq 0$ the first normal stress difference. This representation is more general and independent of any global coordinate system considered. It also involves more physical quantities such as the first normal stress difference, stressline curvatures and derivatives in polymer related directions. \rev{Furthermore, it can provide an interesting framework to interpret flow induced birefringence measurements \cite{fuller1995optical,ober2011spatially} that leverage the optical anisotropy in regions of large stress where polymers are aligned.}  

\rev{Eq.~(\ref{eq:forcing})} can also be seen as an extension of the concept of hoop stress frequently invoked in the context of curved streamlines. The case of curved thin sheets of large polymer stress, as in SAR or EIT, is of particular interest. In this case, $\stressline{1}$ is mostly aligned with the sheet and $\eigval{1} \gg \eigval{2}$, so that at first order, the polymer body force component normal to the sheet is directly proportional to the polymer stress $\eigval{1}$ in the sheet and the curvature $\curvature{1}$ of the sheet. The stresslines $\stressline{1}$ are illustrated in Fig.~\ref{fig:stagnation_points} by the green lines. One can clearly see the alignment between the stresslines, the streamlines and the sheets in regions of large polymer extension, demonstrating that the stretched polymers are mostly aligned with the sheets. \rev{In regions of low polymer extension (e.g., in the recirculation region), there is no alignment between stresslines and streamlines.} 

\begin{figure*}
    \includegraphics[width=\textwidth]{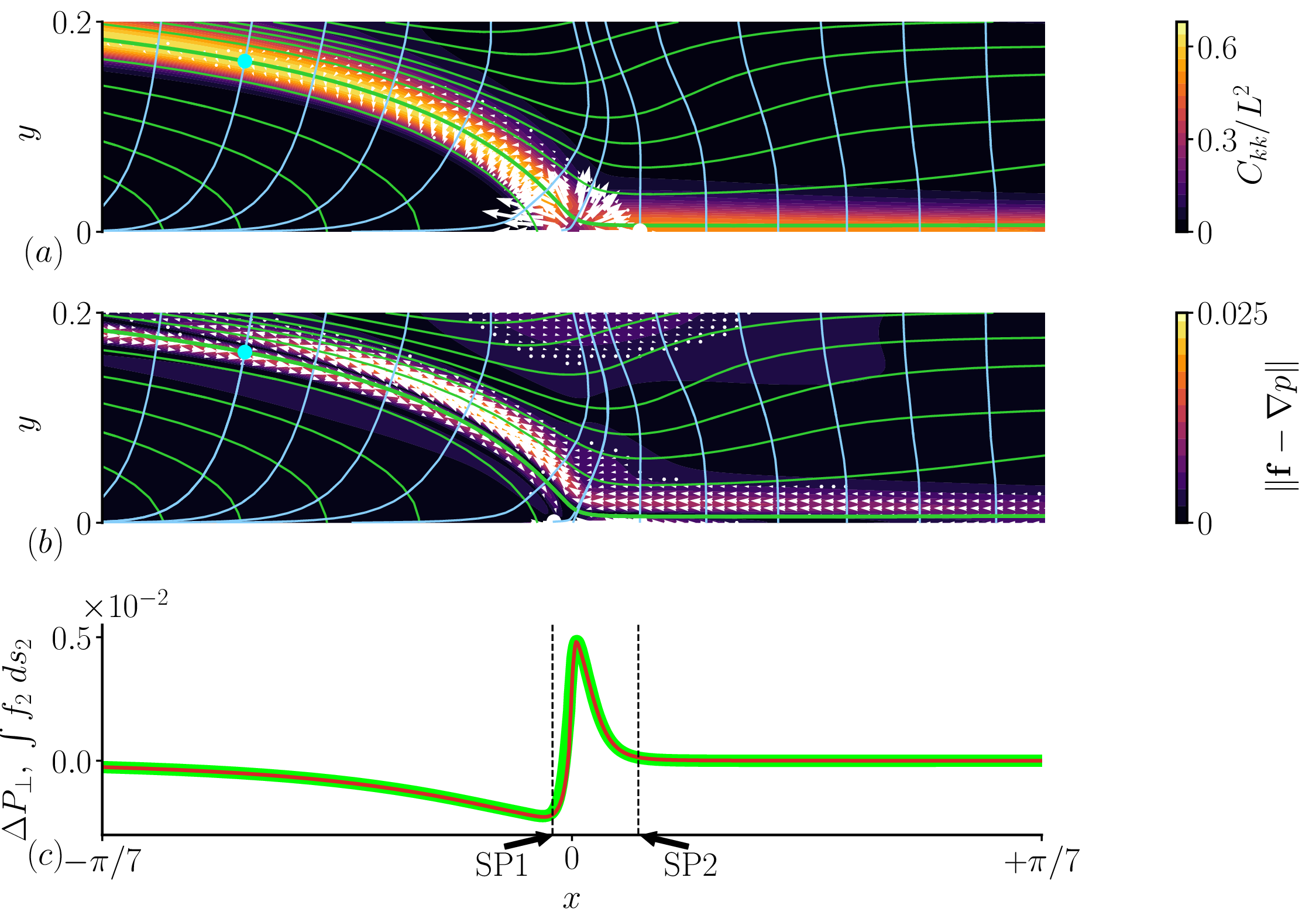}
    \caption{\label{fig:forcing_curvature} \rev{(a):} Scaled polymer body force $\forcingVec$ (arrows) on top of the contour of the normalized polymer extension $C_{kk}/L^2$. \rev{(b):} Scaled resultant forcing $\forcingVec-\boldsymbol{\nabla}p$ (arrows) on top of the contour of its magnitude. \rev{In (a-b): green and blue lines are the stresslines generated from the eigenvectors $\mathbf{e}^{(1)}$ and $\mathbf{e}^{(2)}$, respectively. Note the green stressline passing through the point of maximum of extension $C_{kk}/L^2$ (blue dot), referred to as $\mathcal{S}^{(1)}_\mathrm{max}$ in (c). Vector fields are represented by white arrows (arrows for which the magnitude of the vector field is less than $4\%$ of the maximum are not shown for clarity). (c): Comparison between the pressure jump $\Delta P_\perp=p_\mathrm{II} - p_\mathrm{I}$ (green line) across a stress polymer sheet centered on $\mathcal{S}^{(1)}$ and the polymer body force component integrated over the thickness of the sheet $\int_{-\delta}^{+\delta}f_2~\mathrm{d}s_2$ (red line). The evolutions of both quantities are calculated along the stressline $\mathcal{S}^{(1)}_\mathrm{max}$ and ploted as a function of the horizontal location $x$ for direct correspondence with (a-b). The thickness of the sheet is $2\delta$, $\delta=0.02$, a choice explained in the text.} \rev{$Re=1000$, $Wi=50$, $\beta=0.9$, $L=90$ and $Sc=500$.}}
\end{figure*}

The polymer body force $\forcingVec$ is depicted in Fig.~\ref{fig:forcing_curvature}(a) along with the two families of stresslines. One can identify two specific regions where $\forcingVec$ is significant: along the curved sheet and around the arrowhead tip (pressure minimum). The component $\forcing{2}$ dominates along the curved sheet due to the sheet curvature and large first normal stress difference. The polymer body force thus points primarily towards the center of curvature of the sheet. This force is in turn for the most part counterbalanced by the pressure gradient, explaining the ``bullet shape'' structure of the pressure field shown in Fig.~\ref{fig:ckk_p_u}(b). At the arrowhead tip, the polymer body force shows a radial pattern around the minimum of pressure and polymer extension (between SP1 and SP2). Here again the body force is dominated by the contribution $\curvature{1}\firstNormalStressDiff$ and $\dirDeriv{1}\eigval{1}$, the former dominating in the left and upper right octants and the latter in the right and upper left octants around the pressure minimum. In this region the polymer body force has a very strong divergent contribution, that is again counterbalanced by the pressure gradient (low pressure zone in Fig.~\ref{fig:ckk_p_u}(b)) to ensure incompressibility. 

As just illustrated, both velocity and pressure fields are impacted by the polymer body force, making it difficult to explain the resulting flow from the knowledge of $\forcingVec$ only. In the Stokes flow limit ($Re=0$), it can be easily shown (at least for a specific set of boundary conditions) that the pressure response directly arises from the divergent part and the velocity response from the rotational part of $\forcingVec$. Unfortunately, the Helmholtz-Hodge decomposition of the polymer body force into a divergent and a rotational field is not unique. Moreover, this interpretation is only approximate in the inertial case considered here. Thus, it is impossible to isolate {\em a priori} (that is, knowing only $\forcingVec$) the contributions of $\forcingVec$ that act on the pressure or the velocity field, respectively. Nonetheless, an {\em a posteriori} analysis of the term $\forcingVec - \boldsymbol{\nabla} p$ provides the contribution of the polymer body force to the velocity field. This is shown in Fig.~\ref{fig:forcing_curvature}(b). The largest magnitude of $\forcingVec - \boldsymbol{\nabla} p$ is about $17$ times smaller than that of $\forcingVec$, demonstrating that polymers have their strongest impact on the pressure field. It is also interesting to note in Fig.~\ref{fig:forcing_curvature}(b) that the curved sheet appears to be composed of two adjacent layers in which $\forcingVec - \boldsymbol{\nabla} p$ has opposite directions. These two layers are also visible, but to a much lesser extent, at the base of the spike. Analysis shows that $\dirDeriv{1}\eigval{1} > \curvature{2}\firstNormalStressDiff$ (i.e. $\forcing{1} > 0$) in the upper layer. Moreover, $\dirDeriv{1}\eigval{1}$ is smaller and $\curvature{2}\firstNormalStressDiff$ larger in the lower layer than in the upper layer, such that $\dirDeriv{1}\eigval{1} < \curvature{2}\firstNormalStressDiff$ (i.e. $\forcing{1} < 0$) there. It could be speculated that the resulting torque contributes to bending the sheet. 

Observing that polymer stress is negligible outside of the thin sheet forming the arrowhead (Figs~\ref{fig:streamlines}b, \ref{fig:stagnation_points}, \ref{fig:forcing_curvature}),  we propose to approximate the polymer stress sheet as a viscoelastic interface of thickness $2\delta$, separating two Newtonian regions noted I and II for the inside and outside of the arrowhead, respectively. Such an approximation is inspired from previous works \cite{Rabin1986,Rallison1988}. The observable alignment of the stresslines with the polymer sheet allows for a description of the jump condition of our proposed interface in the frame of reference of the stresslines. The Newtonian stress jump across the interface is therefore derived from Eq.~\eqref{eq:forcing} in a formulation akin to surface tension \cite{Harlen1990} (see supplemental materials for the detailed derivation):
\rev{
\begin{align}
        &\underbrace{p_\domainIndexTwo-p_\domainIndexOne}_{\Delta P_\perp} + 2\frac{\beta}{Re}(\eigvecVec{2}\cdot\mathbf{E}_\domainIndexOne\cdot\eigvecVec{2} - \eigvecVec{2}\cdot\mathbf{E}_\domainIndexTwo\cdot\eigvecVec{2}) \approx \int_{-\delta}^{+\delta}\underbrace{\frac{1-\beta}{\text{Re}} \left(\dirDeriv{2}\eigval{2} + \curvature{1}\firstNormalStressDiff\right)}_{f_2}~\mathrm{d}s_2\,, \label{eq:jumpcond1}\\
        &2\frac{\beta}{Re}(\eigvecVec{1}\cdot\mathbf{E}_\domainIndexOne\cdot\eigvecVec{2} - \eigvecVec{1}\cdot\mathbf{E}_\domainIndexTwo\cdot\eigvecVec{2}) \approx \int_{-\delta}^{+\delta}  \underbrace{\frac{1-\beta}{Re}\left(\dirDeriv{1}\eigval{1} - \curvature{2}\firstNormalStressDiff\right)}_{f_1}~\mathrm{d}s_2\,, \label{eq:jumpcond2}
\end{align}} where $\mathbf{E} = \frac{1}{2}\left(\boldsymbol{\nabla}\mathbf{u} + \boldsymbol{\nabla}\mathbf{u}^\text{T}\right)$ is the strain rate tensor, $\thickness$ is a measure of the half-hickness of the polymer sheet (see Sec.~III of the Supplemental Material \cite{supplemental}). In the curved region of the sheet (head of the arrowhead), the jump in strain rate \rev{(second term of the LHS of Eq.~\ref{eq:jumpcond1}) } and the term $\tilde{\partial}\tau_2$ are found to be small, or even negligible. 

 \rev{Fig.~\ref{fig:forcing_curvature}c shows the evolution of the pressure jump on the LHS of Eq.~\eqref{eq:jumpcond1} and the RHS of the same equation along the stressline that passes through the point of maximum extension shown in Figs.~\ref{fig:forcing_curvature}(a-b). This stressline follows closely the maximum polymer extension in most cross-sections of the arrowhead sheet. The thickness of the sheet in any cross section is defined as the width between the two points where polymer extension is half the maximum polymer extension in that section. An analysis of the sheet shows that this thickness is  $2\delta=0.04$ ($\pm0.004$) in the region of interest, the curved part of the arrowhead (see Sec.~III of the Supplemental Material \cite{supplemental}). For context, the maximum root-mean-squared of pressure in planes parallel to the wall is about $10^{-3}$. The pressure jump $\Delta P_\perp=p_\mathrm{II} - p_\mathrm{I}$ shown in Fig.~\ref{fig:forcing_curvature}c is seven times higher. An analysis of the integral of the forcing term shows that the main contribution comes from the curvature term, especially in the region of high stressline curvature between SP1 and SP2, \textit{i.e.} $\int_{-\delta}^{+\delta}f_2~\mathrm{d}s_2\approx(1-\beta)/Re\int_{-\delta}^{+\delta}\kappa_1N_1~\mathrm{d}s_2$. The excellent agreement between the pressure jump  and the integral of the forcing term across the sheet demonstrates the relation between pressure jump and the curvature of sheets of high polymer stress. } 
 
 The forcing in the direction of the sheet \rev{has the potential to} induces a jump in the shear rate, similarly to the Marangoni effect, as shown in Eq.~\eqref{eq:jumpcond2}. Note that the effect of the two layers with opposite tangential forcing seen in Fig.~\ref{fig:forcing_curvature}(b) could be modeled through a surface torque distribution, which would additionally induce a discontinuity in the tangential velocity across the interface. \rev{The present flow is not found to produce large enough principal polymer stress variation $\dirDeriv{1}\eigval{1}$ along the stressline to create an observable shear rate jump across the sheet. The validation of Eq.~\eqref{eq:jumpcond2} will require future studies in different flows, especially flows with significant $\dirDeriv{1}\eigval{1}$, like plumes of convection cells (see for instance Fig.~2 in reference \cite{dubief2020heat} where polymer stress causes significant velocity deficit at the center of the vertical plumes). Eqs.~(\ref{eq:jumpcond1}-\ref{eq:jumpcond2}) illustrate the possible parallels that may exist between EIT-type flows and interfacial flows, however much work is needed to refine the definition of polymer stress sheets, their thickness, and the regimes where each equation is a satisfactory approximation of the dynamics. }

\textit{Conclusion.} We propose a novel approach to study flow-polymer interaction following the stresslines and highlighting the role of the curvature of the sheets of large polymer extension. The specific case of interest is the steady arrowhead in a two-dimensional periodic channel flow, which eases the analysis step as this structure is steady in a frame of reference moving at constant velocity. This approach makes it possible to analyze in detail the interaction between the polymers and the flow, which is much more challenging to do in chaotic flows where the use of statistical tools may typically hide some interesting features. Although only one set of parameters is considered, this method helps understand and characterize in detail the coherent structures of the polymer stress, pressure, and velocity fields of the steady arrowhead solution. Two different flow regions are highlighted in the moving frame of reference: the recirculation zone and the external flow, both separated by stagnation points. These stagnation points are indicative of extensional flows, which significantly stretch the polymers. We show by means of the newly proposed decomposition along stresslines that the sheets curvature as well as the variation of the stress along the sheets induce a polymer body force influencing both the velocity and the pressure fields, the latter mainly acting in regions where the polymer forcing is convergent/divergent so as to maintain the flow divergent-free. Finally, we derive an approximation by considering the impact of the polymers to be localized on infinitesimal sheets and where the rest of the fluid is seen as Newtonian. Similarly to surface tension, velocity and stress jump conditions can be derived at the interface highlighting how the polymers can typically affect the flow and pressure fields. In particular, it is shown that the pressure jump can be mostly explained by the polymer body force component normal to the sheets (dominated by the curvature term in regions of large anisotropy). Furthermore, the component of the forcing parallel to the sheets induces a jump in shear rate, similar to the Marangoni effect, and possibly a jump in tangential velocity. 

 \textit{Acknowledgments.} This work was supported by the Belgian Fonds de la Recherche Scientifique - FNRS under Grant(s) No PDR T.0216.23. The present research benefited from computational resources made available on Lucia, the Tier-1 supercomputer of the Walloon Region, infrastructure funded by the Walloon Region under the grant agreement n°1910247, and by the Consortium des Équipements de Calcul Intensif (CÉCI), funded by the Fonds de la Recherche Scientifique de Belgique (F.R.S.-FNRS) under Grant No. 2.5020.11 and by the Walloon Region.


\bibliography{references}

\end{document}